\documentclass[prl,twocolumn]{revtex4}

\usepackage{graphicx}

\begin{document}

\title{Adatom incorporation and step crossing at the edges of 2D nanoislands}

\author{Sergey N. Filimonov}
\email{filimon@phys.tsu.ru} \affiliation{Department of Physics,
Tomsk State University, 634050 Tomsk, Russia}
\author{Yuri Yu. Hervieu}
\email{hervieu@elefot.tsu.ru} \affiliation{Department of Physics,
Tomsk State University, 634050 Tomsk, Russia}

\date{\today}

\begin{abstract}
Adatom incorporation into the ``faceted'' steps bordering the 2D
nanoislands is analyzed. The step permeability and incorporation
coefficients are derived for some typical growth situations. It is
shown that the step consisting of equivalent straight segments can
be permeable even in the case of fast egde migration if there
exist factors delaying creation of new kinks. The step consisting
of alternating rough and straight segments may be permeable if
there is no adatom transport between neighboring segments through
the corner diffusion.

\end{abstract}

\maketitle

%\section{Introduction}

Surface evolution in epitaxial growth is known to be highly
sensitive to details of interaction of adatoms with the monoatomic
steps.\cite{Pimpinelli} In the continuous approach such details
should be taken into account in the incorporation and step
permeability coefficients appearing in the boundary conditions for
the surface diffusion equation. However, it is still a common
praxis to treat the kinetic coefficients as the phenomenological
Arrhenius-like constants. One can show that in many cases,
especially when the adatom incorporation into the step is limited
by the process of creation of kinks at the step edge
\cite{Voronkov,Caflisch}, such a simplified approach is not
correct.

Recently we have proposed a method to \emph{derive} the
incorporation and step permeability coefficients of vicinal steps
aligned along high symmetry directions.\cite{Filimonov1} The aim
of the present paper is to extend this approach to construct the
kinetics coefficients of the edges of 2D nanoislands. The
specific shape and small size of the 2D islands give rise to the
peculiarities in the kinetics of adatom incorporation and crossing
the island edge as compared to the case of the ``infinite''
vicinal steps.  So, the length of the edge of a 2D island may be
less than or comparable with the average distance between
neighboring kinks at the edge of a vicinal step situated at the
same surface under the same growth conditions. Then it is highly
probable for the 2D island to be of a strongly polygonized shape,
i.e. to be bordered by a ``faceted'' step containing few or even
no kinks. This is exactly what was observed e.g. in the scanning
tunnelling microscopy studies of growth of Si and Ge on the
Si(111) and (001) surfaces. \cite{Voigtlaender}

The description of adatom incorporation into the ``faceted''
islands inevitably involves analysis of two processes - creation
of kinks at the step ``facets'' and the ``interfacet'' material
transport. Both processes have been addressed in the literature
only in the limiting cases of irreversible attachment of the
terrace adatoms to the edge or weak migration of the edge
adatoms.\cite{Voronkov,Chernov,Michely} In the present paper we
get rid of those simplifying assumptions and derive the
incorporation and step permeability coefficients for two typical
growth situations - the 2D islands with equivalent step segments
and the 2D islands with alternating atomically straight and rough
step segments (as, e.g. the Si islands on the Si(111)-7x7 and
Si(001)-2x1 surfaces, respectively).

\begin{figure}[b]
\includegraphics [angle=0,width=7.5 cm]{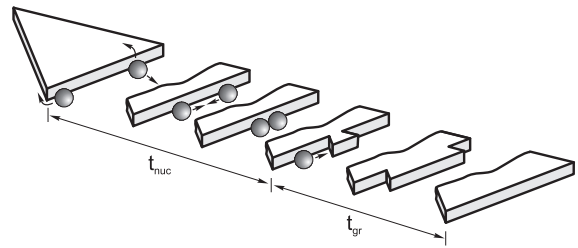}
\caption{Schematic of atomic processes at the edge of the 2D
island during the row-by-row growth process.}
\end{figure}

%\section {The Model}
We will consider propagation of a step segment which length $L$ is
less than the average distance between kinks $L_k$ at the infinite
step considered at the same growth conditions. An adatom attached
to such a segment has four possibilities (Fig. 1): (1) to detach
from the segment back to the terrace from which it came or to the
adjacent terrace (the latter means the crossing of the step); (2)
to leave the segment by rounding the island corners; (3) to meet
another adatom or an unstable cluster at the same segment and in
that way to take part in the 1D nucleation process. After
appearance of the 1D island the adatom can (4) incorporate into
one of two kinks at its ends.

As the result, the row-by-row growth process sketched in Fig.~1
will take place. It includes appearance of the 1D island at the
straight (without kinks) step segment after the expectation time
$t_{nuc}$ and its spreading along the step edge during the mean
time $t_{gr}$. It is essential that no other 1D islands appear
during the time $t_{gr}$. As the crystalline row along the step
segment completes, the 1D nucleation and formation of the new
crystalline row start again.

If the adatoms do not migrate along the step edge then
$t_{nuc}=1/LJ$, where $J$ is the 1D nucleation rate per atomic
site at the step.\cite{Voronkov,Chernov} In this case $t_{nuc}\gg
t_{gr}$ and the maximum of the step permeability is achieved, the
relevant expressions for the kinetic coefficients can be found in
Ref.~8. In the present paper we are interesting in the opposite
limit of fast edge migration, when the traversal time required for
an adatom to visit all sites at the edge $t_{tr}\sim L^2/D_e$ is
much less than the mean time interval between subsequent
attachments of the terrace adatoms to the edge $\Delta
t=1/L(k^+_{le}n_l+k^+_{ue}n_u)$ and the mean time
$t_{res}=1/(k^-_{el}+k^-_{eu})$ that an adatom spends at the
straight step edge before detachment, where $D_e$ is the edge
diffusion coefficient, $k^+_{l(u)e}$ and $k^-_{el(u)}$ are the
attachment ($+$) and detachment ($-$) rate constants, and $n_{l}$
and $n_{u}$ are the concentrations of adatoms on the lower ($l$)
and upper ($u$) terraces in the close vicinity to the 2D island
edge. Bearing in mind comparatively low growth temperatures we
neglect detachment of atoms embedded into the straight step as
well as into the kink and corner sites.

%\section {The Exchange Rate}

Using the above picture of elementary processes we calculate the
net flux (the exchange rate) of adatoms between the island edge
and the adatom gas at the lower terrace averaged over the edge
length and the period $t_{1D}=t_{nuc}+t_{gr}$ of formation of the
crystalline row along the edge
\begin{equation}
        g_l=\frac{1}{L}\int_0^{L}\frac{1}{t_{1D}}\int_0^{t_{1D}}j_l(x,t)dxdt.
        \label{meanexchange}
\end{equation}
Here $x$ is the coordinate along the island edge and $j_l(x,t)$ is
the local net flux. In the limit of fast edge migration an adatom
which attaches to the edge segment during the time interval
$t_{gr}$ has no chances to detach from this segment (it may
however to leave the segment via corner rounding). Then one can
express $j_l(x,t)$ as
\[j_l(x,t)=\left\{
\begin{array}{rl}
k^+_{le}n_l-k^-_{el}n_e(x), & \mbox{within the time interval } t_{nuc} \\
k^+_{le}n_l, & \mbox{within the time interval } t_{gr}
\end{array} \right. \]
where we assume that the concentration of adatoms on the terrace
does not change considerably during the time interval $t_{1D}$
(the concentrations of the terrace adatoms $n_{l}$ and $n_{u}$ are
considered as unknown variables for which the usual quasi
steady-state approximation holds).

The integration gives
\begin{equation}
        g_l=(1-\tau_k)(k^+_{le}n_l-k^-_{el}\bar{n}_e)+\tau_kk^+_{le}n_l
        \label{meangeneral}
        \end{equation}
where $\tau_k=t_{gr}/(t_{gr}+t_{nuc})$ is the fraction of time
when every adatom  attaching to the edge contributes to the
formation of the crystalline layer along one of the edge segments
and
\[\bar{n}_e=\frac{1}{L}\int_0^{L}n_e(x)dx\]
is the mean concentration of the edge adatoms.  The concentration
$n_e(x)$ is found as the solution of the edge diffusion equation
\begin{equation}
D_e \frac{d^2
n_e}{dx^2}-(k^-_{el}+k^-_{eu})n_e(x)+k^+_{le}n_l+k^+_{ue}n_u=0
\label{diffeq}
\end{equation}
with the boundary conditions describing incorporation of the edge
adatoms at the kink sites or/and leaving the edge segment at the
island corners. In the following we summarize our major results
for some typical growth situations.

%\section {The Kinetic Coefficients}

%\subsection{The 2D islands with equivalent step segments}

\textbf{The 2D islands with equivalent step segments.} Assuming
fast migration of the edge adatoms around the island corners (this
process has a high probability e.g., in the case of triangular 2D
Si islands on the Si(111)-7x7 surface)\cite{Voigtlaender} we
adopt the periodical boundary conditions at the corners. In this
case the concentration of the edge adatoms does not depernd on
$x$ and is given by
$n_e=(k^+_{le}n_l+k^+_{ue}n_u)/(k^-_{el}+k^-_{eu})$ . Substituting
this expression into Eq.~(\ref{meangeneral}) we obtain the
exchange rate $g_{l}$ in the standard form\cite{Ozdemir}
\begin{equation}
g_{l}=\nu_{l}(n_{l}-\tilde{n})+\nu_p(n_{l}-n_{u}). \label{gl1}
\end{equation}
The first term in the right part of Eq.~(\ref{gl1}) is the flux of
adatoms incorporating into the kinks at the island edge and the
second term is the flux of adatoms crossing the edge without
visiting the kinks, $\nu_l$ and $\nu_p$ are the incorporation and
step permeability coefficients, respectively. The coefficients are
given by
\begin{equation}
\nu_{l}=\tau_kk^+_{le}; \qquad
\nu_p=\frac{(1-\tau_k)k^+_{le}k^-_{eu}}{k^-_{eu}+k^-_{el}}.
\label{coeff1}
\end{equation}
Similar expression for the flux $g_u$ of adatoms leaving the upper
terrace and relevant kinetic coefficients may be obtained by the
substitution $l$ for $u$ and \textit{vice versa} in
Eqs.~(\ref{gl1}) and (\ref{coeff1}).

The ability of the adatoms from the lower terrace to cross the
edge and thus climb up the 2D island top may be characterized by
the ratio
\begin{equation}
\eta_l=\frac{\nu_p}{\nu_l}=
\frac{t_{nuc}k^-_{eu}}{t_{gr}(k^-_{eu}+k^-_{el})}. \label{etaeq}
\end{equation}
As can be seen from Eq.~(\ref{etaeq}), the edge segment may be
permeable ($\eta_l\gg 1$) if its propagation is limited by the
kink creation ($t_{gr}\ll t_{nuc}$) even if migration of the edge
adatoms is fast.

%\subsection{The 2D islands with inequivalent step segments}

\textbf{The 2D islands with inequivalent step segments.} We have
considered the case of the 2D island with alternating atomically
straight and rough edge segments as e.g., SA and SB edge segments
of the rectangular 2D Si islands on the Si(100)-2x1 surface. Here
the probabilities for an adatom to cross the straight segment or
incorporate into it are both affected by the ability of the edge
adatom to travel around the step corners. Assuming that the
adatom does not return back from the neighboring rough segments
we get
\begin{equation}
g_{l}=\nu_{l}(n_{l}-\tilde{n})+\nu_p(n_{l}-n_{u})+\nu_{ec}n_{l},
\label{gl2}
\end{equation}
where the term $\nu_{ec}n_{l}$ is the flux of adatoms attaching to
the segment and leaving it via the corner rounding. The kinetic
coefficients appearing in Eq.~(\ref{gl2}) are written in the form
\[
\nu_{l}=\kappa\tau_kk^+_{le};
\]
\[\nu_p=\frac{(1-\tau_k)[1-f_c(q_L)]k^+_{le}k^-_{eu}}
{k^-_{eu}+k^-_{el}};
\]
\[\nu_{ec}=[(1-\kappa)\tau_k+(1-\tau_k)f_c(q_L)]k^+_{le},
\]
where
\[
f_c(q_L)=\frac{\tanh(q_L)}{q_L\left[1+2q_L\tanh(q_L)D_e/(k_{ec}L)\right]}
\]
is the probability that an adatom, attached to the step segment
containing no kinks, will leave the segment via the corner
rounding before detachment, $q_L$ is the ratio of the segment
length to the mean length of the adatom migration along the
infinite step, $k_{ec}$ is the rate constant for corner rounding
and $\kappa$ ($0.5\leq \kappa\leq 1$) is the probability that the
edge adatom will find the kink when the latter is present at the
given step segment.

The permeability ratio in this case is given by
\[
\eta_l=\frac{\nu_p}{\nu_l+\nu_{ec}}=\frac{(1-\tau_k)[1-f_c(q_L)]k^-_{eu}}
        {[\tau_k+(1-\tau_k)f_c(q_L)](k^-_{el}+k^-_{eu})}.
\]
Here the neighboring step segments act as a pair of kinks settled
at a short ($L<L_k$) distance. This diminishes crossing the step
by the terrace adatoms. Our calculations give that the step
segment may be permeable only if the energy barrier for the corner
rounding is greater than the barrier $E_e$ for the edge migration
by $\Delta E_{ec}>E^-_{es}-E_e-k_BT\ln(L/2)$, where $E^-_{es}$ is
the smallest from the barriers for detachment to the upper and
lower terraces.

%\section {Summary and Discussion}

In conclusion, we have derived the incorporation and step
permeability coefficients for two typical growth situations
involving the 2D islands bounded by the ``faceted'' steps.  It has
been shown that adatom incorporation into such islands has some
peculiarities which are reflected by the structure of the kinetic
coefficients.

It should be noted that in spite of the linear form of
Eqs.~(\ref{gl1}) and (\ref{gl2}) the exchange rates are in general
the \emph{non-linear} functions of the adatom concentrations in
the vicinity of the island edge because propagation of the edge
segments involves a non-linear process of formation of kinks by
the non-equilibrium 1D nucleation mechanism. The coefficients
$\nu_l$ and $\nu_p$ are, in fact, the functions of the adatom
concentrations $n_u$ and $n_l$, and they reduce to the
Arrhenius-like constants only when the system is close to
equilibrium or in the case of irreversible attachment of the
adatoms to the island edge (i.e. when $\tau_k=1$).

To apply our model to the particular system of interest one needs
to specify the characteristic time scales $t_{gr}$ and $t_{nuc}$.
Evidently, $t_{gr}\sim 1/(k^+_{le}n_l+k^+_{ue}n_u)$ with the
coefficient of proportionalty (order unity) depending on the 2D
island geometry and intensity of the corner rounding processes.
The nucleation time $t_{nuc}$ is the inverse of the nucleation
rate which can be calculated with the statistical theory of the
island-on-island nucleation.\cite{Krug} An application of the
outlined strategy to modeling of the mass-transport during the
formation of the multilayer Ge nanoislands on Si(111)\cite{Teys}
can be found elsewhere.\cite{Filimonov3}

\section*{Acknowledgments}

This work has been supported by INTAS (03-51-5015) and RFBR
(03-02-17644). S.F. is grateful to the Alexander von Humboldt
Foundation for support.

\end{document}